\documentclass[aps,prb,showpacs,twocolumn,10pt,floatfix]{revtex4-1}    
\usepackage{amssymb,graphicx,color,amsmath,bm} 
\usepackage[all]{xy}
\usepackage[bookmarks, colorlinks=true, breaklinks]{hyperref} 
\hypersetup{linkcolor=blue,citecolor=blue,filecolor=black,urlcolor=blue}

\begin{document}

\title{Electronic Structure of Silicon-Based Nanostructures}

\author{G. G. Guzm\'{a}n-Verri$^{1,2}$ and L. C. Lew Yan Voon$^{1}$}
\affiliation{$^1$Department of Physics, Wright State University, 3640 Colonel Glenn Highway, 
Dayton, Ohio 45435, USA \\
$^2$Centro de Investigaci\'{o}n en Ciencia e Ingenier\'{i}a de Materiales, 
Universidad de Costa Rica, 2060 San Jos\'{e}, Costa Rica}

\date{\today}

\begin{abstract}
We have developed an unifying tight-binding Hamiltonian that can account for the electronic properties 
of recently proposed Si-based nanostructures, namely, Si graphene-like sheets and Si nanotubes.  
We considered the $sp^3s^*$ and $sp^{3}$ models up to first- and  second-nearest neighbors, respectively. 
Our results show that the Si graphene-like sheets considered here are metals or zero-gap semiconductors, 
and that the corresponding Si nanotubes follow the so-called Hamada's rule [Phys. Rev. Lett. {\bf 68}, 1579 1992]. Comparison to a recent
{\it ab initio} calculation is made.
\end{abstract}


\maketitle


\section{Introduction}

After the first synthesis of carbon nanotubes (CNT's) by Iijima more than a decade ago,~\cite{Iijima91a}
other types of nanotubes have been predicted and experimentally observed such
as GaN, BN, and AlN among others.~\cite{Goldberger03a,JZhang05a,Xi06a,Zhukovskii06a}
However, it was not until fairly recently that the most obvious alternative candidate
for creating graphene-like sheets and nanotubes was proposed: Si.~\cite{Fagan00a}
In addition, different
structures have been proposed, each one with different hybridizations: 
$sp^{2}, sp^{2}-sp^{3}$ and $sp^{3}$. So far, most studies 
agree that the $sp^{2}$ configuration is the least 
favorable one and, in contrast, the $sp^{3}$ configuration is
one of the most favorable ones because of stability 
reasons.~\cite{Fagan00a, Yang05a, RQZhang05a, Seifert01a}

On the experimental side, there are now at least six independent reports of 
the fabrication of silicon nanotubes in the 
laboratory.~\cite{Sha02a, Jeong03a, Chen05a, Tang05a, Crescenzi05a, Castrucci06a}
On the theoretical side, only a handful of papers have explored the 
electronic properties of these nanomaterials;
moreover, practically all of them correspond to {\it ab initio} 
calculations.~\cite{Fagan00a, Yang05a, RQZhang05a, Seifert01a, Yan06a}
The above early work was reviewed by Perepichka and Rosei.~\cite{Perepichka06a}

In the present work, we apply tight-binding (TB) models, so successfully used
to study the electronic properties of graphene and CNT's,~\cite{Saito98a}
to the Si nanostructures that have $sp^{2}$ and $sp^{3}$ hybridization.  We will
refer to these structures as silicene, Si~($111$), Si hexagonal nanotubes (Si h-NT's), and
Si gear-like nanotubes (Si g-NT's). Silicene is a two dimensional sheet 
with a honeycomb lattice of lattice constant $a$ made 
out of Si atoms which have $sp^{2}$ hybridization.
Thus, silicene has the same structure as
a graphene sheet but it is composed of Si atoms instead of C atoms. 
A Si~($111$) layer has a lattice structure which is the same as the honeycomb lattice for silicene, 
except that
one set of atoms (e.g., B) is vertically displaced (e.g., down) from the
A-plane due to the $sp^3$ bonding (see Fig.~\ref{fig:Si-111-sheet}). 
Single-walled Si h-NT's and Si g-NT's are formed by rolling up, respectively, 
a silicene and a Si~($111$) sheet. 

Our goal is two-fold. First, we would
like to compare the electronic properties of silicene and Si~($111$) to 
graphene, and the electronic properties of Si h-NT's and Si g-NT's to CNT's.
Second, inconsistencies between an {\it ab initio} calculation and a proposed
$\pi$-TB model, both by Yang and Ni~\cite{Yang05a} 
motivated us in developing a coupled $\sigma$--$\pi$
TB model. Results on the effect of the coupling will be presented.

The present paper is organized as follows: in Section II 
we describe the TB theory for the Si nanostructures, 
in Section III we present and discuss
our results, and in Section IV we give 
the conclusions. An appendix
has been added for further details about the model 
in question.


\section{The tight-binding model}
We now present a new theory of the band structure of a single Si sheet.
This theory applies to both the flat Si sheet and the Si~($111$) layer.

\begin{figure*}
\centering
\includegraphics[width=12cm]{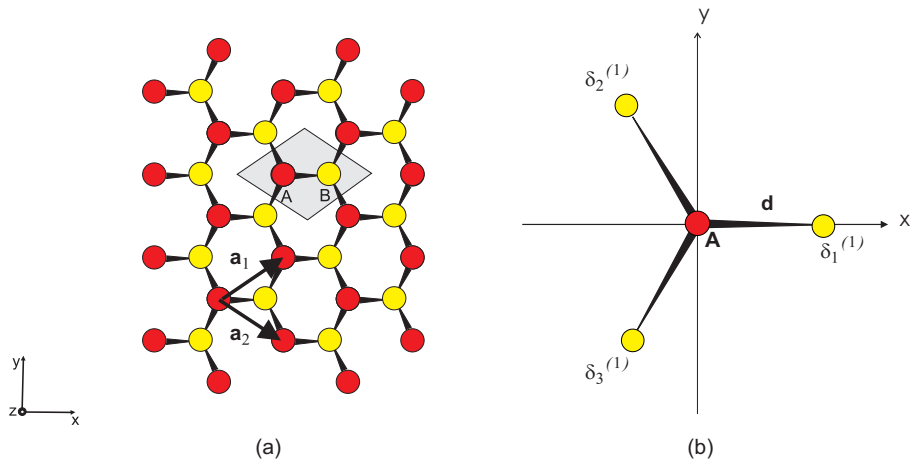}
\caption{(Color online) Lattice of Si~($111$). (a) Two-dimensional representation of a Si~($111$) sheet. 
The atoms labeled as A are all in the $xy$ plane ($z_A=0$) and all the B atoms 
are located below the plane $\left[z=-a/(2\sqrt{6})\right]$. Hence, the sheet is composed by two atomic planes: one of A atoms and another of B atoms. The A plane is above the B plane. Notice that the $z$ 
axis points towards the reader. The vectors ${\bf a}_{1}$ and ${\bf a}_{2}$ are the two-dimensional 
basis vectors and the shaded area is the Si~($111$) unit cell. 
(b) First-nearest neighbors of Si~($111$).}
\label{fig:Si-111-sheet}
\end{figure*}

The reason a unifying Hamiltonian is possible is because the lattice structure 
of a Si~($111$) sheet is similar to graphene. 
 Figure~\ref{fig:Si-111-sheet}(a) is a two-dimensional
representation of the lattice in question. 
The A atoms are in the $xy$-plane and the B atoms are out of it and located
at $z=-a/(2\sqrt{6})$ (the $z$ axis points towards the reader). 
Thus, the word sheet in the present work means one atomic plane 
of A atoms above one atomic plane of B atoms, {\it i.e.}, a 
sheet is two atomic planes. The basis vectors of the lattice are 
${\bf a}_{1}=(a/2)\left(\sqrt{3},-1\right)$ 
and ${\bf a}_{2}=(a/2)\left(\sqrt{3},1\right)$ with magnitude $a$.~\cite{Saito98a} 
If we label as $l$ the Si-Si bond distance, then $a=l/\sqrt{2}$. 
We will see that the choice of this coordinate system
facilitates the transition from the Si~($111$) sheet to the silicene one.   
The shaded area corresponds to the two-dimensional unit cell of 
Si~($111$). Notice that the basis vectors and the unit
cell of Si~($111$) are equal to the ones in the honeycomb lattice of graphene.~\cite{Saito98a}
Due to this similarity with graphene, the Brillouin zone of the Si~($111$) lattice 
is the same as the one of graphene.    

In order to compute the band structure of Si~($111$), we use a 
first-nearest neighbor (1NN) $sp^{3}s^{*}$ and a second-nearest neighbor (2NN) $sp^3$ 
orthogonal tight-binding model. For these two models, we derive 
their respective $10 \times 10$ and $8 \times 8$ Hamiltonians. These choices were dictated by the
availability of good Si TB parameters which correspond to 
Vogl et al.~\cite{Vogl83a}~($sp^{3}s^{*}$) and to Grosso and Piermarocchi
~\cite{Grosso95a}($sp^{3}$). The authors are aware of newer TB parametrizations,~\cite{Klimeck00a, Martins05a} however, these parametrizations do not reproduce well the Si bulk bands along the $K\Gamma$ direction, which is important for the Si nanostructures. 

Like silicene and
graphene, the wavevector ${\bf k}$ of Si~($111$) in the Hamiltonian is
two dimensional, however, their dispersion functions $g_{j}({\bf k})$, differ due 
to the $\sigma-\pi$ coupling (see Appendix A). Furthermore, we use the two-center approximation~(TCA) 
in
order to obtain the angular dependence in the TB parameters. For the sake of completeness, the TCA
parameters are reproduced in Table~\ref{t:TCA-parameters}.
\begin{table}
\caption{Silicon two-center parameters obtained from 
Vogl et al.~\cite{Vogl83a} and Grosso and Piermarocchi.~\cite{Grosso95a} Blank spaces
correspond to parameters that do not belong to the model.}
\centering
 \label{t:TCA-parameters}
\begin{ruledtabular}
\begin{tabular}{lrr} 
Parameter             &   Vogl et al.~\cite{Vogl83a} & Grosso and Piermarocchi.~\cite{Grosso95a}  \\ \hline
$E_{s}$               &  $-4.2000$  &   $-4.0497$  \\
$E_{p}$               &    $1.7150$ &   $1.0297$  \\
$E_{s^{*}}$           &    $6.6850$ &    \\ 
$(ss\sigma)_{1}^{AB}$ &  $-2.0750$  & $-2.0662$\\
$(sp\sigma)_{1}^{AB}$ &    $2.4808$ & $2.0850$ \\
$(pp\sigma)_{1}^{AB}$ &    $2.7163$ &  $3.1837$\\
$(pp\pi)_{1}^{AB}$    &  $-0.7150$  & $-0.9488$ \\
$(s^{*}p\sigma)_{AB}^{1}$ & $2.3274$&     \\
$(ss\sigma)_{2}^{AA}$ &             &  $0.0000$\\
$(sp\sigma)_{2}^{AA}$ &             &  $0.0000$\\
$(pp\sigma)_{2}^{AA}$ &             & $0.8900$\\
$(pp\pi)_{2}^{AA}$    &             &  $-0.3612$\\
\end{tabular}
\end{ruledtabular}
\end{table}  
We first specify the position vectors of the  1NN for Si~($111$):
\begin{eqnarray}
\label{eq:Si(111)_1NN}
\mbox{\boldmath{$\delta$}}_{1}^{(1)}&=&\left(\frac{a}{\sqrt{3}}, 0, -\frac{a}{2 \sqrt{6}}\right),\nonumber\\
\mbox{\boldmath{$\delta$}}_{2}^{(1)}&=&\left( -\frac{a}{2 \sqrt{3}},\frac{a}{2},-\frac{a}{2 \sqrt{6}}\right),\\
\mbox{\boldmath{$\delta$}}_{3}^{(1)}&=&\left(-\frac{a}{2 \sqrt{3}},  -\frac{a}{2}, -\frac{a}{2 \sqrt{6}}\right),\nonumber
\end{eqnarray}
and for the 2NN:
\begin{align}
\label{eq:Si(111)_2NN}
\mbox{\boldmath{$\delta$}}_{1}^{(2)}&=\left(0,a,0\right),
& \mbox{\boldmath{$\delta$}}_{2}^{(2)}&=\left(0,-a,0\right),\nonumber\\
\mbox{\boldmath{$\delta$}}_{3}^{(2)}&=\left(\frac{a \sqrt{3}}{ 2 },
                                     -\frac{a}{2},0\right),
& \mbox{\boldmath{$\delta$}}_{4}^{(2)}&=\left(-\frac{a \sqrt{3}}{2},\frac{a}{2},0\right),\\
\mbox{\boldmath{$\delta$}}_{5}^{(2)}&=\left(\frac{a \sqrt{3}}{2},\frac{a}{2},0\right),
& \mbox{\boldmath{$\delta$}}_{6}^{(2)}&=\left(-\frac{a \sqrt{3}}{ 2 },
                                  -\frac{a}{2},0\right). \nonumber
\end{align}
For simplicity, Fig.~\ref{fig:Si-111-sheet}(b) only shows 
the position vectors of the 1NN. Notice that
the $x$- and $y$-components of the 1NN in Si~($111$) correspond to the ones of graphene. The
non-zero $z$-components are the vertical displacements of the $B$ atoms of the 
Si~($111$) sheet. In other words, the choice
of the coordinate system facilitates the transition from Si~($111$) and silicene by 
making the $z$ component of the position
vectors in Eq.~(\ref{eq:Si(111)_1NN}) equal to 
zero. The 2NN coincide for both sheets. 

The dispersion $E({\bf k})$ for the Si~($111$) sheet is shown in
Figs.~\ref{fig:Si-sheets}(d) and \ref{fig:Si-sheets}(e). For the $sp^{3}$ model, 
we can find analytic formulae for the dispersion relation at the $\Gamma$ point,
\begin{eqnarray}
E_{p+}(\Gamma) =  E_{p}&+&3[(pp\sigma)_{2}^{AA}+(pp\pi)_{2}^{AA}]\nonumber\\
        &+& \frac{1}{3}[4(pp\sigma)_{1}^{AB}+5(pp\pi)_{1}^{AB}], \\
E_{p-}(\Gamma) =  E_{p}&+&3[(pp\sigma)_{2}^{AA}+(pp\pi)_{2}^{AA}]\nonumber\\
        &-& \frac{1}{3}[4(pp\sigma)_{1}^{AB}+5(pp\pi)_{1}^{AB}],\nonumber 
\end{eqnarray}
where $E_{p\pm}$ is twofold degenerate. 

Next, we explain how this formalism developed
for Si~($111$) can be transfered to silicene.

The $\pi$ and $\sigma$ bands in silicene can be obtained from the 
previous Hamiltonian by making the $z$-component 
of the 1NN equal to zero and substituting the 
appropriate direct cosines of the 1NN position vectors. 
As a result, the $\pi$ and $\sigma$ bands are decoupled, as in graphene.
This allow us to consider the bands independently.

For the $\pi$ bands, we recover the well known $2 \times 2$ Hamiltonian but now
including 2NN interactions,
\begin{equation}\label{eq:H-sil}
H({\bf k})=\begin{bmatrix}E_{p}+(pp\pi)_{2}^{AA}g_{25}({\bf k})  & \gamma_{0} g_{12}({\bf k})\\
                     \gamma_{0} g_{12}^{*}({\bf k}) & E_{p}+(pp\pi)_{2}^{AA}g_{25}({\bf k})
\end{bmatrix},
\end{equation}
where $\gamma_{0}(=|(pp\pi)_{1}^{AB}|)$ is the transfer integral and the 
$g_{j}({\bf k})$ functions are given in Eq.~(\ref{eq:g-func-Si-111}).  
The energy dispersion relation for silicene is found from
Eq.~(\ref{eq:H-sil}),
\begin{equation}\label{eq:bs-pi-silicene}
E({\bf k})=E_{p}+(pp\pi)_{2}^{AA}g_{25}({\bf k})\pm(pp\pi)_{1}^{AB}w({\bf k}),
\end{equation}
where the dispersion function 
$w({\bf k})\equiv |g_{12}({\bf k})|
=\sqrt{1+4\cos\frac{\sqrt{3}k_{x}a}{2}\cos\frac{k_{y}a}{2}+4\cos^{2}\frac{k_{y}a}{2}}$. 
Notice that if we make the 2NN contribution equal to zero, {\it i.e.}, $(pp\pi)_{2}^{AA}=0$,
we recover graphene's energy dispersion in the 1NN approximation.~\cite{Saito98a}

For the $\sigma$ bands, the matrix elements of the Hamiltonian
correspond to similar ones given in Eq.~(\ref{eq:H-sub-matrices}) for
Si~($111$), but now the $g_{j}({\bf k})$ functions correspond to
the ones given in Eq.~(\ref{eq:g-func-graph}). These bands are shown
in Figs.~\ref{fig:Si-sheets}(a) and~\ref{fig:Si-sheets}(b).

By evaluating the Hamiltonian at the $\Gamma$ point, we can obtain analytical
formulae for the dispersion relations,
\begin{eqnarray*}
E_{p_{z}\pm}(\Gamma) = E_{p} &\pm& 3(pp\pi)_{1}^{AB},\\
E_{s\pm}(\Gamma) = E_{s} &+& 6(ss\sigma)_{2}^{AA} \pm 3(ss\sigma)_{1}^{AB},\\
E_{p\pm}(\Gamma) = E_{p} &+& 3\left[(pp\sigma)_{2}^{AA}+(pp\pi)_{2}^{AA}\right]\\
                   &\pm&  \frac{3}{2}\left[(pp\pi)_{1}^{AB}+(pp\sigma)_{1}^{AB}\right],
\end{eqnarray*}
where  $E_{p\pm}(\Gamma)$ is twofold degenerate.

Finally, the Si h-NT and Si g-NT band structures
are computed from Si~($111$) and silicene by 
imposing boundary conditions on ${\bf k}$
 along the chiral direction.~\cite{Saito98a}
The results are shown in Fig.~\ref{fig:Si-NTs}. 

We now proceed to examine our findings obtained from
the TB theory. 


\section{Results and discussion}
\begin{figure}
\includegraphics[width=8.5cm, height=8cm]{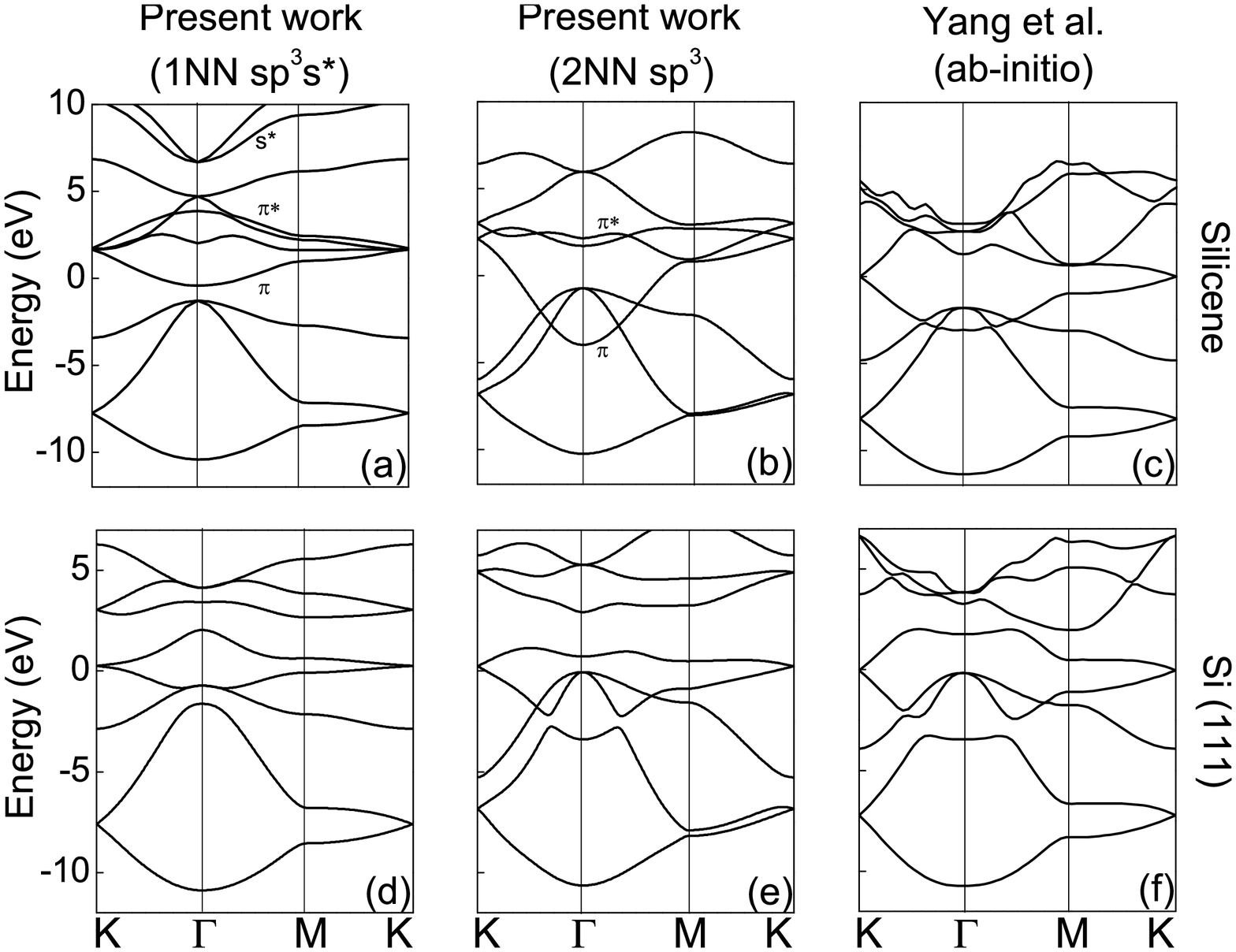}
\caption{Band structure of silicene and of Si~($111$) obtained from
our TB models compared to the {\it ab-initio} results from 
Yang and Ni.~\cite{Yang05a}}
\label{fig:Si-sheets}
\end{figure}

\begin{figure}
\includegraphics[width=9cm, height=9cm]{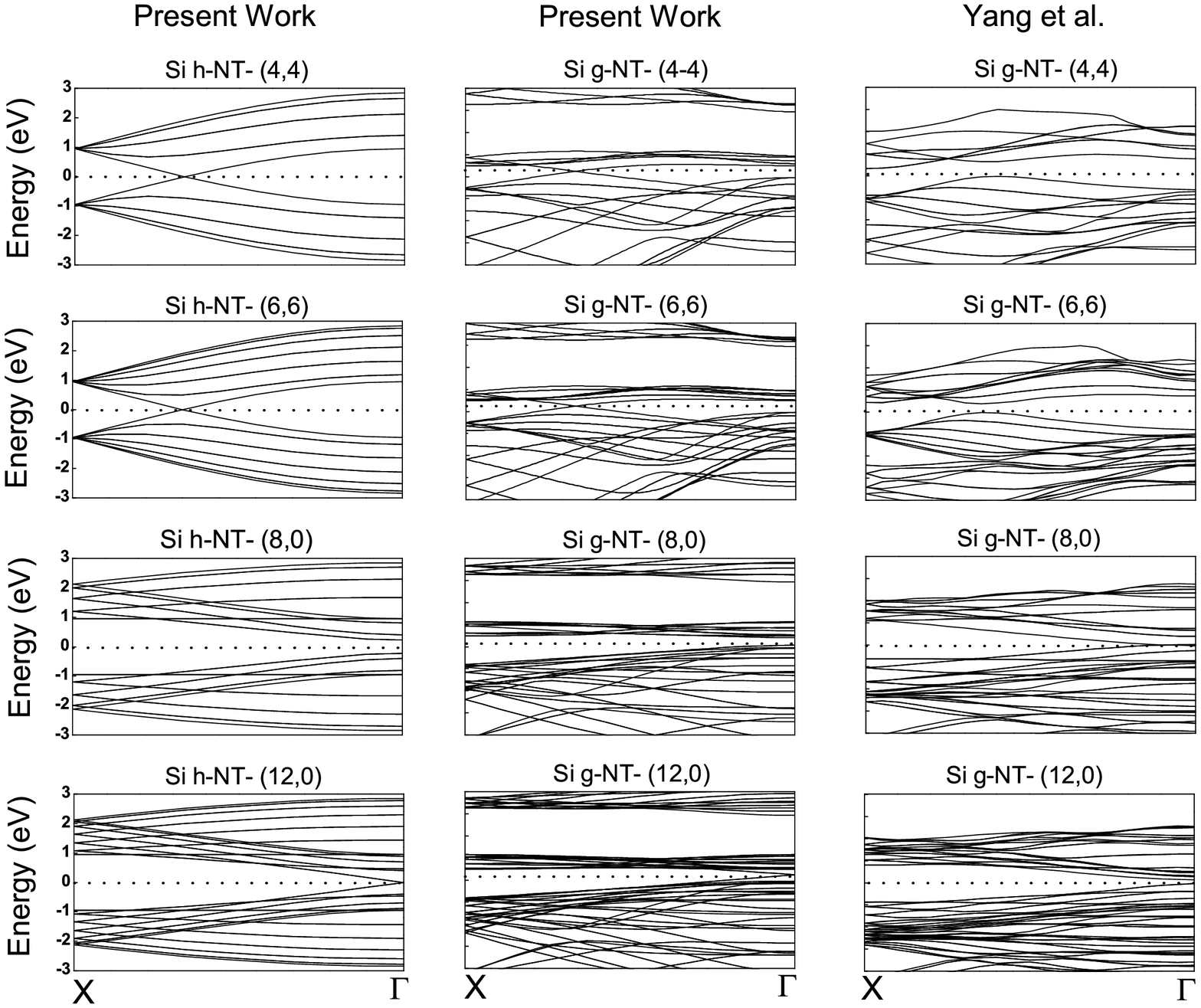}
\caption{Band structures of  Si h- and Si g-NT's according to our calculations and 
to Yang and Ni.~\cite{Yang05a} The two-center parameters used in our calculations 
were taken from Grosso and Piermarocchi
.~\cite{Grosso95a}}
\label{fig:Si-NTs}
\end{figure}

 The electronic band structures for the Si sheets and the Si-NT's are
shown in Figs.~\ref{fig:Si-sheets} and \ref{fig:Si-NTs}. Here, we 
discuss them separately.
  
\subsection{Silicene}

The silicene band structure is shown in 
Figs.~\ref{fig:Si-sheets}(a) and~\ref{fig:Si-sheets}(b) according to the $sp^{3}s^*$
and $sp^3$ models, respectively.
Like in graphene, the silicene $\pi$ bands are not coupled to the $\sigma$ bands due 
to the planar and orbital symmetries.~\cite{Saito98a}
When we compare the $\pi$ energy bands of silicene from the $sp^{3}s^{*}$ model to the corresponding
ones in graphene, we notice
that they have a similar form. This occurs because both sheets have the same 
lattice structure. As far as the $\sigma$ bands are concerned, the valence bands in 
silicene 
have been 
lowered down, therefore, 
the crossings that occur in graphene between the $\pi$ and $\sigma$ bands 
do not occur in silicene. Different results are obtained when we perform the same comparison 
using the $sp^{3}$ model. 
The $\pi$ band maintains the form as in graphene, nevertheless,
the $\pi^*$ band changes its curvature as it approaches $\Gamma$ from the $K\Gamma$ and the $M\Gamma$ 
directions. This
change could occur because of the 2NN interactions or because of the signs of the TB parameters. In order 
to determine which is the case, one has to perform further calculations on the curvature of the band, 
which will be presented
somewhere else. In contrast to the $sp^{3}s^{*}$ model, the $\pi$ and $\sigma$ valence bands do cross. 
This shows how sensitive the band structure is to
the TB parametrization of the model. 

It is known that close to the $K$ point, graphene shows a linear dispersion,~\cite{Novoselov05a,Berger06a,Ando05a,Peres06a}
\begin{equation}\label{eq:E=mv2}
E(p)= \pm v_{0}|p|,
\end{equation}  
where $v_{0}$ and $p(=\hbar k)$ are the Fermi velocity and the carrier momentum, respectively.
In graphene, it has been measured that $v_{0}\approx 10^{6}$~m/s.~\cite{Novoselov05a, Berger06a}
The linearity of $E(p)$ is responsible for the electrons to behave as Dirac massless fermions.~\cite{Novoselov05a,Ando05a}
Figures~\ref{fig:Si-sheets}(a) and~\ref{fig:Si-sheets}(b) show that silicene has
a linear dispersion close to the $K$ point. By performing a linear fitting,
we find that the Fermi velocity $v_{0}$ in Eq.~(\ref{eq:E=mv2}) is of the order of $10^{5}$~m/s for
both models. Compared to graphene, the electrons move slower in 
silicene. This occurs since the $\pi$ bonds, which are the responsible for conduction in the sheets, 
are weaker in silicene
than in graphene~(the atomic distance in silicene is greater than in graphene). 

Next, let us now compare our $sp^{3}s^{*}$  results to Yang's {\it ab-initio} ones. 
The band structure of silicene obtained by Yang and Ni~\cite{Yang05a}
is shown in Fig.~\ref{fig:Si-sheets}(c). When compared 
one notices certain differences. For example, in the neighborhood of 
the $\Gamma$ point in Fig~\ref{fig:Si-sheets}(a), both conduction and valence bands have the
opposite curvature with respect to Fig.~\ref{fig:Si-sheets}(c). 
Possibly, this difference can be attributed to the 1NN approximation in the
$sp^{3}s^*$ model. There is agreement in the silicene $\pi$ and the 
lower $\sigma$ bands, however, this
is not the case for the upper $\pi^{*}$ band, which has opposite curvature. Moreover,
there are no crossings between the $\pi$ and $\sigma$ valence bands. As it was 
discussed previously, this is due to the TB parameters.

\subsection{Si~($111$)}

\begin{figure}
\begin{center}
\includegraphics[width=8.6cm, height=3.6cm]{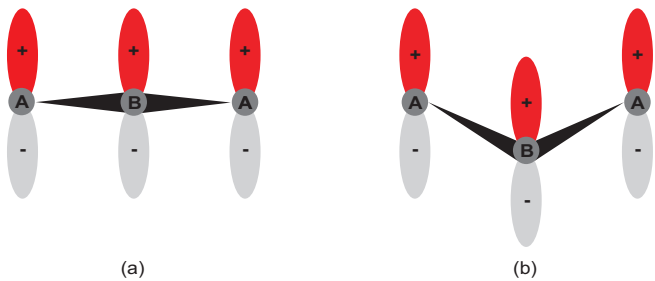}
\caption{(Color online) The $sp^{2}$ and $sp^{3}$ hybridizations in (a) silicene and in (b) Si~($111$).} 
\label{fig:sp2-vs-sp3}
\end{center}
\end{figure}

The band structure of Si~($111$) is shown in Fig.~\ref{fig:Si-sheets}(d)
and~\ref{fig:Si-sheets}(e) 
for the $sp^{3}s^{*}$ and the $sp^{3}$ models, respectively. 
In Si~($111$), the $sp^{3}$ hybridization causes  
a coupling between the $\pi$ and the $\sigma$ states. The effect of it 
is evident in the anti-crossing of the originally uncoupled bands in 
graphene.~\cite{Saito98a}
Note, however, that 
the $\pi$ band is still doubly degenerate
at the $K$ point. In  graphene, the degeneracy at this point is required by the
hexagonal symmetry.~\cite{Milosevic98a} When we lower the B atoms in silicene to create the 
Si~($111$) sheet, as it is shown in Fig.~\ref{fig:sp2-vs-sp3}, this symmetry is not removed hence,
the $\pi$ bands are still degenerate at  $K$ in Si~($111$).

Notice that the $sp^{3}$ and the $sp^3s^{*}$ models
lead to different bands, particularly, close to the $\Gamma$ point. As the bands approach 
this point, the curvature of some of them changes. Consider, for instance, the second-lowest
valence band in Figs.~\ref{fig:Si-sheets}(d) and~\ref{fig:Si-sheets}(e). Close to $\Gamma$, this band
has negative curvature in the former figure, while it has positive curvature in the latter one. 
These differences might be due to 2NN interactions. We computed the band structure of Si~($111$) 
using two parameter
sets for a $sp^{3}$ TB model by Chadi and Cohen;~\cite{Chadi75a} one of them for 1NN only, and the other one
for 1NN plus one 2NN interaction. These results are not presented here, however, we mention
that the band without the 2NN interaction has negative curvature, like Fig.~\ref{fig:Si-sheets}(d), 
and the band with only
one 2NN interaction has positive curvature, like Fig.~\ref{fig:Si-sheets}(e). Another possible reason
for the curvature
differences is the sign of the TB parameters. Further calculations on the curvature of the bands at the 
$\Gamma$ point are needed in order to determine if it is due to the 2NN interactions or the TB parameters.
These calculations will be shown somewhere else.

As far as the eigenstates are concerned, they are different in graphene and in Si~($111$). The
$\pi$ bands in both structures are a good example of this. Whereas the eigenstates 
of the the $\pi$ bands in graphene are $p_{z}$ orbitals, the eigenstates of the $\pi$ bands
in Si~($111$) correspond to a linear combination of $s, p_{x}, p_{y}$ and $p_{z}$ orbitals. This
occurs because in Si~($111$), the $\pi$ and $\sigma$ bands are coupled.

Notice in Figs.~\ref{fig:Si-sheets}(d) and~\ref{fig:Si-sheets}(e) that close to the $K$ point, the dispersion
relation is linear, which indicates the presence of Dirac massless 
fermions. By performing a linear fitting, we find that the Fermi velocity
$v_{0}$ is of the order of $10^{4}\,$m/s. Compared to graphene,
 the Dirac fermions move slower in the Si~($111$) sheet. 
In order to understand the difference in velocities, let us look at the 
hybridizations of both sheets, as it is shown in Fig~\ref{fig:sp2-vs-sp3}.

On the one hand, 
each atom shows $sp^{2}$ hybridization in graphene. In this hybridization, each atom has a lobe that 
is perpendicular
to the sheet plane~(the sheet corresponds to the $xy$-plane and the lobe 
 is oriented along the positive
$z$-axis as it is shown in Fig.~\ref{fig:sp2-vs-sp3}(a)). All lobes are oriented along the same direction, i.e., the $z$-axis, and, therefore,
they form $\pi$ bonds with their 1NN. The $\pi$ bonds are responsible for the conducting character
of the sheet.

On the other hand, Si~($111$) shows $sp^3$ hybridization. Figure~\ref{fig:sp2-vs-sp3}(b) shows that
each atom has a lobe that is
perpendicular to the sheet, however, neighboring atoms have their lobes pointing in 
opposite directions, i.e., the positive and negative $z$-axes. Due to the alternating
orientation, a lobe that points along the, say, positive $z$-axis, will not form a $\pi$ bonding
with its 1NN, but with its 2NN. 

Since the $\pi$ bonds in graphene occur between 1NN, the coupling is stronger than 
in Si~($111$), where the bonds occur between 2NN. An electron finds it ``easier'' to
tunnel from one atom to the another one when the coupling is stronger. This explains why
electrons in graphene move faster than they do in Si~($111$).           

The band structure of the Si~($111$) sheet computed by Yang and Ni.~\cite{Yang05a} is
shown in Fig.~\ref{fig:Si-sheets}(f).
When compared to our $sp^{3}$ findings, 
we find good agreement between them, especially, along the
$K \Gamma M$ directions. Close to $\Gamma$, the curvature
problem is overcame by including some TCA interactions from
2NN in the Hamiltonian, namely, $(pp\sigma)_{2}^{AA}$ and 
$(pp\pi)_{2}^{AA}$. Most of the differences occur along the $MK$ direction, 
for instance,
our calculations show that the second-lowest valence band should have
the opposite curvature when it is compared to Yang and Ni~\cite{Yang05a}
The positive curvature of this band comes from the strong repulsion
induced by the $p$-like band located above it.     

\subsection{Si h-NT}
Nanotubes are fully characterized by their chiral vector 
${\bf C}_{h}=n{\bf a}_{1}+m{\bf a}_{2}$, 
where $n$ and $m$ are integer numbers, and 
${\bf a}_{1}$ and ${\bf a}_{2}$
are given above.~\cite{Saito98a}
The band structures of Si h-NT's with chiral
 vectors $(4,4),(6,6),(8,0)$ and $(12,0)$ are shown 
in the left-most column of Fig.~\ref{fig:Si-NTs}. The dashed line
corresponds to the Fermi level. These
bands were obtained by substituting the quantized nanotube wave vector
${\bf k}$ in Eq.~(\ref{eq:bs-pi-silicene}). In this figure, the dispersion $g_{25}({\bf k})$,
introduced by the 2NN interactions, its hardly noticeable, therefore,
the band structure of Si h-NT's looks very similar to the one of
CNT's in the 1NN approximation. It is due to this similarity, that we
neglect the dispersion $g_{25}({\bf k})$ in the band structure of Si h-NT's 
and discuss it as if it were a 1NN approximation only. In the 1NN case, 
the only difference between the CNT and the Si h-NT band structures is 
a scaling factor,
 which corresponds to the transfer integral
 $\gamma_{0}$~($\gamma_{0}=-3.033$~eV for C and $\gamma_{0}=-0.949$~eV for Si).
 Whether or not a Si h-NT would be conductor or semiconductor,
 does not depend on $\gamma_{0}$ but on its symmetries, thus,
 Si h-NT will have similar electronic properties as CNT's, {\it i.e.},
 if $n-m$ is a multiple of $3$, the nanotube is a metal, otherwise,
 it is a semiconductor. We will refer to this property as Hamada's rule.~\cite{Hamada92a}
From Hamada's rule, we conclude that
 all armchair Si h-NT's are conductors~($n=m$) and zig zag tubes are conductors
 if $n$ is a multiple of $3$~($m=0$). The band structures for Si h-NT's shown
 in Fig.~\ref{fig:Si-NTs} confirm this rule. 

\begin{figure}
\begin{center}
\includegraphics[width=7cm, height=5cm]{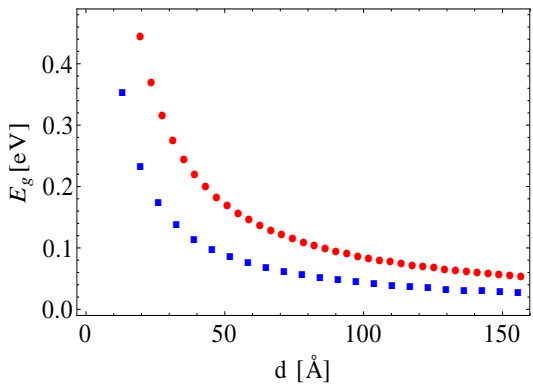}
\caption{(Color online) Band gap of CNT's~(circles) and Si h-NT's~(squares) as a function of their diameters.}
\label{f:EgvsD} 
\end{center}
\end{figure}

We proceed to compare the energy band gaps between CNT's and Si h-NT's. According to 
Dresselhaus et al.~\cite{Dresselhaus94a} the band gap is given by 
\begin{equation}\label{eq:NT-bandgap}
E_{g}=2\gamma_{0}\mbox{Min}\,w\left(k\frac{{\bf K}_{2}}{K_{2}}
                         +\frac{1}{3}(N \pm 1){\bf K}_{1}\right),
\end{equation}
where ${\bf K}_{1}$ and ${\bf K}_{2}$ are the reciprocal vectors of graphene
and silicene,
 $\gamma_{0}$ is the transfer integral,
$N$ is the number of hexagons in the nanotube unit cell,
 $k$ is the wave vector, and Min$\,w$ is 
the minimum of the graphene and silicene
 dispersion relation $w(=|g_{12}({\bf k})|)$ with respect to $k$. In particular,
 for zig zag tubes, 
the minimum occurs at $k=0$,~\cite{Saito98a}
according to Eq.~(\ref{eq:NT-bandgap}), this yields to,
\begin{equation}
E_{g}=2\gamma_{0} \sqrt{1 + 4\cos{\frac{\pi q_{0}}{n}}+4\cos^{2}{\frac{\pi q_{0}}{n}}},
\end{equation}
where $q_{0}=(1/3)(N \pm 1)$ and $N=2n$. In Fig.~\ref{f:EgvsD}, we show our 
results for the band gap of both CNT's and h-SiNT's as a function of the diameter. 
For a given diameter, the band gap $E_{g}$ of Si h-NT's is smaller than the band gap 
of CNT's. In order to understand this result, we should look to the approximate 
expression for $E_{g}$ derived by Saito et al.~\cite{Saito98a} and Ando,~\cite{Ando05a}
 \begin{equation}
E_{g}\sim\frac{|\gamma|}{d}, 
\end{equation}
where
$\gamma = \sqrt{3}a \gamma_{0}/2$ and $d=|{\bf C}_{h}|/\pi$ 
(the nanotube diameter). For a given diameter, $E_{g}$ depends only 
on the parameter $\gamma$, which depends on the transfer integral~(recall that 
the transfer integral corresponds to the interaction between neighboring $p_{z}$ orbitals).
Since this interaction is smaller for Si h-NT's than for CNT's, the band gap is 
greater for the latter.

Notice that the $1/d$ dependence in the nanotube band gap is expected. If we let $d\rightarrow \infty$,
then $E_{g}\rightarrow 0$. The limit $d\rightarrow \infty$ corresponds to making a nanotube
 thicker and thicker, and, therefore, more similar to graphene, which has a zero band gap.

We proceed to compute and to compare the effective masses of Si h-NT's and CNT's. 
By differentiating the band structure of zig zag nanotubes given in 
Eq.~(\ref{eq:bs-pi-silicene}), and neglecting the 2NN interactions, we obtain  
an analytical formula for their effective masses at the $K$ point,
\begin{eqnarray}\label{eq:NT-effmass}
m_{\pm}^{*}&=&{\left. \frac{\hbar^{2}}{d^{2}E/dk^{2}}\right|_{k=0}}\nonumber\\
           &=&
\pm\frac{2\hbar^{2} }{3 \gamma_{0} a^{2}}
\frac{\sqrt{1+4\cos{\frac{\pi q_{0}}{n}}+4\cos^{2}{\frac{\pi q_{0}}{n}}}}
{\cos{\frac{\pi q_{0}}{n}}}.
\end{eqnarray}

\begin{figure}
\begin{center}
\includegraphics[width=7cm, height=5cm]{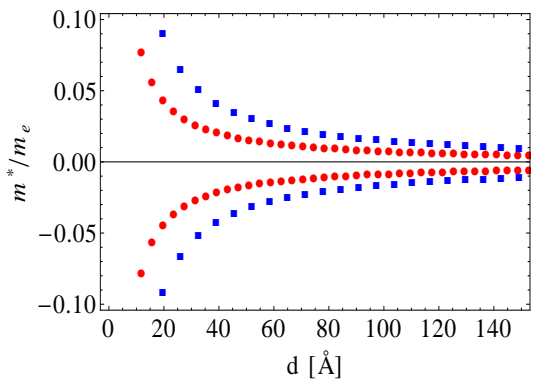} 
\caption{(Color online) Effective masses of CNT's~(circles) and Si h-NT's~(squared) as a function of their diameters.}
\label{fig:EMvsD} 
\end{center}
\end{figure}

Equation~(\ref{eq:NT-effmass}) is plotted in Fig.~\ref{fig:EMvsD}. First, notice the
general trend for Si h-NT's and CNT's that the effective masses decrease as  
their diameter increases. This occurs because electrons in a Si h-NT~(CNT) with large
diameter, behave more similar to electrons in silicene~(graphene). In 
other words, if we increase the nanotube diameter, we will see that the electrons start
 ``losing''
their masses, until they become Dirac massless fermions. Second, notice that for a given
diameter, electrons in Si h-NT's are heavier, in magnitude, than electrons in CNT's. The
top valence band and lowest conduction band are flatter than the corresponding ones in CNT's,
which explains their mass difference.

We point out that the effective masses can be obtained from ${\bf k}\cdot{\bf p}$ theory as well.
Ando~\cite{Ando05a} derived the following dispersion relation for the nanotube band gap,   
\begin{equation}{\label{eq:kp}}
E_{n\pm}(k)=\pm \gamma\sqrt{\kappa(n)^{2}+k'^2},
\end{equation}
with
\[
\kappa = \frac{2\pi}{L}\left(n-\frac{\nu}{3}\right) ,n=0,\pm1,\pm2,...
\]
where $k'$ is the wave vector measured from the $K$ point, $L$ is the magnitude of the 
chiral vector and 
$\vert \nu \vert = 1$ for semiconducting nanotubes. Hence, the effective mass of the zig zag
tubes is
\begin{equation}{\label{eq:kpem}}
m_{\pm}^{*}=\left. \frac{\hbar^{2}}{d^{2}E_{0\pm}/dk'^{2}}\right|_{k'=0}
     =\pm \frac{4 \hbar^{2}}{3\sqrt{3}\gamma_{0} a}\left(\frac{1}{d}\right).
\end{equation}
This equation confirms what was stated above about the nanotube effective masses using 
TB theory, that is, $m^{*}$ decreases as the $d$ increases since it 
explicitly depends upon $d^{-1}$.

\subsection{Si g-NT's}

The second column of Fig.~\ref{fig:Si-NTs} shows the band structures 
of Si g-NT's with
chiral vectors $(4,4),(6,6),(8,0)$ and $(12,0)$ obtained from the $sp^{3}$ model.  
Notice that for all graphs but the ones
in the second column, the Fermi energy is at zero. From this figure,
one notices that there is a proliferation of the number of bands in the Si g-NT's
when compared to CNT's. In CNT's there is a total of $2N$ bands.~\cite{Saito98a} The
factor of $2$ comes from the $2 \times 2$ Hamiltonian, and $N$ bands come
from the quantization of the wave vector ${\bf k}$ along the ${\bf K}_{1}$
direction. In Si g-NT's, 
there are $8N$ bands; eight from the $8 \times 8$
Hamiltonian, and the other $N$ from the same quantization in ${\bf k }$. Hence,
Si g-NT's have four times the number of bands that CNT's have. Moreover, there
are 4 electrons per Si atom and a total of $2N$ atoms in the g-NT unit cell, 
hence, there are $4\times 2N = 8N$ electrons per unit cell. In particular,
armchair and zig zag g-NT's, have $16n$ bands and electrons per unit cell~($N=2n$). 
Table~\ref{t:h- vs g-NT's}
summarizes some of the general characteristics of the h-NT and g-NT energy
bands.

\begin{table}[h]
\caption{General characteristics of the h-NT and g-NT energy bands. Here, e$^-$ and u.c. refer to 
electrons and the nanotube unit cell, respectively.}
\label{t:h- vs g-NT's}
\centering
\begin{ruledtabular}
\begin{tabular}{lcc}
          &  h-NT~($sp^{2}$) & g-NT~($sp^{3}$)\\ \hline
e$^{-}$ per atom   & $1$  &  $ 4$ \\
atoms per u.c.   & $2N$  &   $2N$ \\
Number of bands   & $2N$  &   $8N$ \\
e$^{-}$ per u.c. & $2N$ &    $8N$ \\
\end{tabular}
\end{ruledtabular}
\end{table}

After wrapping the Si~($111$) sheet into a
zig zag g-NT, we find that the $(8,0)$ tube is a semiconductor
with a gap of $0.34$~eV, and the $(12,0)$ is either a metal or a semiconductor 
of gap zero.
The armchair $(4,4)$ and $(6,6)$ tubes could be metals or zero gap 
semiconductors, as well. In order to completely characterize the electronic
behavior of the last three nanotubes, we need their density of states~(DOS), which
is not done here. However,
due to the similarities with CNT's, we would expect them to have a finite 
DOS at the Fermi level and thus, to be metals.

In general, Si g-NT's obey Hamada's rule like CNT's do. In order
to understand this, we apply a similar reasoning used for CNT's.~\cite{Saito98a} We first started
with the Si~($111$) band structure~(c.f. Fig.~\ref{fig:Si-sheets}(e)), which has zero gap
at the $K$ point of its Brillouin zone. Second,
we sliced the Si~($111$) bands by quantizing the 
wavevector ${\bf k}$ along the ${\bf K}_{1}$ direction. 
Third, we used the zone-folding
approximation to plot the Si g-NT bands. If any of the ``slices'' cuts
the $K$ point then, the g-NT is a conductor, if it does not, then
it is a semiconductor.  In the language of group theory, the degeneracies
caused by the vertical and the horizontal mirror plane symmetries in CNT's are not
removed in the gear-like tube.   

Next, let us compare our Si g-NT energy bands to the
bands by Yang and Ni.~\cite{Yang05a}
From the previous comparisons of the Si sheets, we find better 
agreement between our calculations and the {\it ab-initio} ones by Yang 
and Ni~\cite{Yang05a} when we use our 2NN $sp^{3}$ TB model.
Hence, we
only compare our energy bands obtained from this theory to the 
{\it ab-initio} ones.

The Si g-NT energy bands by Yang and Ni~\cite{Yang05a} are reproduced
in the left-most column of the Fig~\ref{fig:Si-NTs}. We find that all our band structures differ from
their work. Whereas all our armchair tubes are metals or zero-gap 
semiconductors, their corresponding tubes are semiconductors. 
Our zig-zag tubes follow 
Hamada's rule, while theirs do not. 

The authors argue that in zig zag Si g-NT's with
small diameter like $(8,0)$, gaps do not occur because of the 
$\sigma^{*}-\pi^{*}$ coupling. We think that this hypothesis is
doubtful, since our theory does include this coupling and it does not
close the gap of the small semiconducting tubes like $(8,0)$. Furthermore, band gaps tend to open and not 
to close when couplings
are added. A possible explanation for the discrepancy is that the {\it ab initio}
calculation suffers from the density-functional-theory band-gap problem.

These significant 
differences between the TB and the {\it ab initio} bands are quite surprising 
since both approaches coincide on the band
structure of Si~($111$). We point out, though, that in their report, Yang and Ni~\cite{Yang05a} 
do not 
explain the gap openings when the Si~($111$)
sheet is rolled up to form Si g-NT's.~\cite{Yang05a} One could think that
curvature effects in Yang's results might open a gap in Si g-NT's, nonetheless, such
gap should be, at most, of the order of  meV. Consider a CNT and a
Si g-NT with the same chirality, for instance, ${\bf C}_{h}=(12,0)$.
It is known that the gap
opening $\Delta E$ due to curvature in CNT's, is inversely proportional to it
and it is close to $10$~meV.~\cite{Saito98a, Reich04a} The transfer
integral $|\gamma|$, can be used as an energy scale for the nanotubes, thus,
$\Delta E \sim |\gamma|/d^2$. The ratio between the openings in CNT's and Si g-NT's
is then given by
\begin{eqnarray}\label{eq:gap}
&& \frac{\Delta E_{\rm Si}}{\Delta E_{\rm C}} \sim 
\frac{|\gamma_{\rm Si}|/d_{\rm Si}^2}{|\gamma_{\rm C}| / d_{\rm C}^2}
= \frac{|\gamma_{\rm Si}|}{|\gamma_{\rm C}|}\left(\frac{d_{\rm C}}{d_{\rm Si}}\right)^{2} \nonumber\\
\Rightarrow && \Delta E_{\rm Si} \sim \frac{|\gamma_{\rm Si}|}{|\gamma_{\rm C}|}
               \left(\frac{d_{\rm C}}{d_{\rm Si}}\right)^{2} \Delta E_{\rm C}.
\end{eqnarray}
Consider a C- and a Si-NT with diameters $d_{\rm C}=30\,$\AA~and $d_{\rm Si}=46\,$\AA, 
respectively.~\cite{Yang05a, Saito98a}
Substituting  $|\gamma_{\rm C}|=3.033$~eV, $|\gamma_{\rm Si}|=0.949$~eV and 
$\Delta E_{\rm C} = 10$~meV~\cite{Yang05a, Saito98a} in Eq.~(\ref{eq:gap}), 
we find that
\[
\Delta E_{\rm Si} \approx 1 \mbox{meV}. 
\]  
The band gap in the $(12,0)$ Si g-NT of Yang and Ni~\cite{Yang05a}~(see band structure in
the lower right corner of Fig.~\ref{fig:Si-NTs}) is close to $200$~meV so,
curvature cannot be responsible for the whole opening of the gap in Yang's nanotubes.

Our TB scheme, on the 
other hand, provides an explanation for the gap behavior in all g-NT's openings in terms of
Hamada's rule. For example, our $(12,0)$ Si g-NT has, according to Hamada's rule, 
a zero gap since this corresponds to a 
zig zag tube with $n$ multiple of 3.
      

\section{CONCLUSIONS}

The electronic properties of silicene,
Si~($111$), Si h-NT's and Si g-NT's were studied via a TB approach. 
We derived $sp^{3}s^*$  
and $sp^{3}$ Hamiltonians up to 1NN and 2NN, respectively.
 
We compared the band structures of Si~($111$) and of silicene to the one
of graphene. Since
all of these materials have in-plane symmetry, the $\pi$ bands are 
two-fold degenerate at $K$. We expect Si~($111$) and silicene to be either
semiconductors of band gap zero or metals. Electrons in
the neighborhood of the $K$ point should 
behave as Dirac massless fermions due to the presence of 
the Dirac cone in both structures. However, the Fermi velocities $v_{0}$
in Si~($111$)~($10^{4}$~m/s) and in silicene~($10^{5}$~m/s) are
smaller than the one in graphene~($10^{6}$~m/s).
Electron tunneling between Si atoms is less favorable in Si~($111$) and
silicene than in graphene because the $\pi$ interaction is weaker in the 
the first two cases.

Silicon h-NT's and Si g-NT's were compared to CNT's, as well. The band structure
of Si h-NT's and of CNT's are similar two each other. Even though we performed
calculations including 1NN and 2NN, we found that the effect of the latter ones on the
bands is negligible. This allowed us to make further approximations when
calculating the band gap and effective masses of Si h-NT's. 
In the case of zig zag semiconductor
Si h-NT's, the gap is inversely proportional to the
tube diameter, as in zig zag CNT's, nonetheless, for a given diameter,
Si h-NT's will have a smaller gap. The magnitude of the effective masses 
is also inversely 
proportional to their diameter, however, for a given diameter Si h-NT's
have greater mass, which makes CNT's more suitable for transport properties.
As far as Si g-NT's are concerned, we found that they follow Hamada's rule 
as CNT's do, even though they show different hybridizations.

Our calculations for all the Si-based materials considered here 
were compared to the {\it ab initio} calculations performed by
Yang and Ni.~\cite{Yang05a}
When comparing silicene and Si~($111$), we found that the 2NN $sp^{3}$ 
model is in better agreement with Yang's band structures than  the 
1NN $sp^{3}s^*$ model. For this reason, we
chose the $sp^{3}$ model in order to reproduce the energy dispersions
of Si nanotubes. 

For Si h-NT's, our band structures agree with the ones obtained by
Yang and Ni:~\cite{Yang05a} they all follow Hamada's rule. 
In contrast, they disagree for all Si g-NT's. Whereas our calculations show
that these nanotubes 
also follow this rule, Yang's calculations do not. 
We emphasize that the $\sigma$--$\pi$ coupling  
does not close the gap for Si g-NT's with small diameters, contrary to the
hypothesis of Yang and Ni.~\cite{Yang05a} Our calculations also
show how critical it is to obtain accurate TB parameters for 
applications to Si sheets and nanotubes.
  
\section{ACKNOWLEDGMENTS}
We thank Yang and Ni for providing the data used in
Figs.~\ref{fig:Si-sheets} and~\ref{fig:Si-NTs}.
The work was supported by an NSF CAREER award (NSF Grant No. 0454849),
and by a Research Challenge grant from Wright State University and the Ohio
Board of Regents. 

\appendix
\section{Tight-binding Hamiltonian}

In this appendix, we explicitly give the Hamiltonian of our model. 
The $sp^{3}s^{*}$ Hamiltonian has the form
\begin{equation} 
H({\bf k}) = \begin{bmatrix}
   h_{AA}      & h_{AB} \\
   h_{BA}      & h_{BB}     
   \end{bmatrix}, 
\label{eq:Hamiltonian}
\end{equation} 
where the $h_{ij}$'s are $5 \times 5$ sub-matrices 
which are given 
as follows: 
\begin{widetext}
\begin{eqnarray}\label{eq:H-sub-matrices}
&&{h_{AA}}= 
\begin{bmatrix}
  (s/s)_{AA} &  (s/x)_{AA} &  (s/y)_{AA} &  0 & (s/s^*)_{AA} \\
  &  (x/x)_{AA} &  (x/y)_{AA} &  0 & (x/s^{*})_{AA} \\
  & &  (y/y)_{AA} &  0 & (y/s^{*})_{AA} \\
  &\dagger & &  (z/z)_{AA} & 0 \\
  &&&& (s^{*}/s^{*})_{AA}  
   \end{bmatrix},\nonumber\\
&& h_{AB}   = \begin{bmatrix} 
(s/s)_{AB} & (s/x)_{AB} & (s/y)_{AB} & (s/z)_{AB} & (s/s^*)_{AB}\\
(x/s)_{AB} & (x/x)_{AB} & (x/y)_{AB} & (x/z)_{AB} & (x/s^*)_{AB}\\
(y/s)_{AB} & (y/x)_{AB} & (y/y)_{AB} & (y/z)_{AB} & (y/s^*)_{AB}\\
(z/s)_{AB} & (z/x)_{AB} & (z/y)_{AB} & (z/z)_{AB} & (z/s^*)_{AB}\\
(s^*/s)_{AB} & (s^*/x)_{AB} & (s^*/y)_{AB} & (s^*/z)_{ac} &(s^*/s^*)_{AB}
\end{bmatrix},\\ 
&& h_{BB}   = \begin{bmatrix}
  (s/s)_{AA}^* &  -(s/x)_{AA}^{*} &  -(s/y)_{AA}^* &  0 & (s/s^*)_{AA}^* \\
  &  (x/x)_{AA}^* &  (x/y)_{AA}^* &  0 & (x/s^{*})_{AA}^* \\
  & &  (y/y)_{AA}^* &  0 & (y/s^{*})_{AA}^* \\
  &\dagger & &  (z/z)_{AA}^* & (z/s^{*})_{AA}^* \\
  &&&& (s^{*}/s^{*})_{AA}^*  
   \end{bmatrix},\nonumber
\end{eqnarray}
\end{widetext}
and $h_{BA}=[h_{AB}]^{\dagger}$. In these equations, $h_{AA}$ and $h_{AB}$ correspond
to the interaction between atoms at $A-A$ and $A-B$ lattice points, respectively. In
the case of the $sp^{3}$ Hamiltonian, each sub-matrix is $4 \times 4$ instead.
The matrix elements are given as follows:
\begin{widetext}
\begin{eqnarray*}
(s/s)_{AA} &=& (s/s)_{BB}^{*} = E_{s} + (ss\sigma)_{2}^{AA}g_{13}({\bf k}), \nonumber\\
(s/x)_{AA} &=& -(s/x)_{BB}^{*} = (sp\sigma)_{2}^{AA}g_{14}({\bf k}), \nonumber\\
(s/y)_{AA} &=& -(s/y)_{BB}^{*} = (sp\sigma)_{2}^{AA}g_{15}({\bf k}), \nonumber\\
(s/z)_{AA} &=& -(s/y)_{BB}^{*} = 0,\nonumber\\
(x/x)_{AA} &=& (x/x)_{BB}^{*} = E_{p} + (pp\sigma)_{2}^{AA}g_{16}({\bf k}) 
            + (pp\pi)_{2}^{AA}g_{17}({\bf k}), \nonumber\\
(x/y)_{AA} &=& (x/y)_{BB}^{*} =\left[(pp\sigma)_{2}^{AA}-(pp\pi)_{2}^{AA}\right]g_{18}({\bf k}), \nonumber\\
(x/z)_{AA} &=& (x/z)_{BB}^{*} = 0, \nonumber\\
(y/y)_{AA} &=& (y/y)_{BB}^{*} = E_{p} + (pp\sigma)_{2}^{AA}g_{19}{\bf k}+(pp\pi)_{2}^{AA}g_{20}({\bf k}), \nonumber\\
(y/z)_{AA} &=& (y/z)_{BB}^{*} = 0, \nonumber\\
(z/z)_{AA} &=& (z/z)_{BB}^{*} = E_{p} + (pp\pi)_{2}^{AA}g_{25}({\bf k}),\nonumber\\
(x/x)_{AB} &=& (pp\sigma)_{1}^{AB}g_{3}({\bf k})+ (pp\pi)_{1}^{AB}g_{4}({\bf k}), \nonumber \\
(y/y)_{AB} &=& (pp\sigma)_{1}^{AB}g_{6}({\bf k})+ (pp\pi)_{1}^{AB}g_{4}({\bf k}),\nonumber\\   
(z/z)_{AB} &=&(pp\sigma)_{1}^{AB}g_{11}({\bf k})+ (pp\pi)_{1}^{AB}g_{12}({\bf k}), \nonumber \\  
(s/x)_{AB} &=&-(x/s)_{AB}=(sp\sigma)_{1}^{AB}g_{1}({\bf k}), \nonumber\\
(s/y)_{AB} &=&-(y/s)_{AB} =(sp\sigma)_{1}^{AB}g_{2}({\bf k}), \nonumber\\
(s/z)_{AB} &=&-(z/s)_{AB}=(sp\sigma)_{1}^{AB}g_{8}({\bf k}),\nonumber\\
(x/y)_{AB} &=&(y/x)_{AB} =\left[(pp\sigma)_{1}^{AB}-(pp\pi)_{1}^{AB} \right]g_{5}({\bf k}),\nonumber\\ 
(x/z)_{AB} &=&(z/x)_{AB} =\left[(pp\sigma)_{1}^{AB}-(pp\pi)_{1}^{AB} \right]g_{9}({\bf k}),\\ 
(y/z)_{AB} &=&(z/y)_{AB} = \left[(pp\sigma)_{1}^{AB}-(pp\pi)_{1}^{AB} \right]g_{10}({\bf k}), \nonumber\\
(s^*/s^*)_{AA} &=& (s^*/s^*)_{BB}^{*} = E_{s^*} + (s^*s^*\sigma)_{2}^{AA}g_{13}({\bf k}), \nonumber\\
(s^*/x)_{AA} &=& -(s^*/x)_{BB}^{*} = (s^*p\sigma)_{2}^{AA}g_{14}({\bf k}), \nonumber\\
(s^*/y)_{AA} &=& -(s^*/y)_{BB}^{*} = (s^*p\sigma)_{2}^{AA}g_{15}({\bf k}), \nonumber\\
(s^*/z)_{AA} &=& -(s^*/y)_{BB}^{*} = 0,\nonumber\\
(s/s^*)_{AA} &=& (s/s^*)_{BB}^{*} = (ss^*\sigma)_{2}^{AA}g_{13}({\bf k}), \nonumber\\
(s^*/x)_{AB} &=&-(x/s^*)_{AB}= (s^*p\sigma)_{1}^{AB}g_{1}({\bf k}), \nonumber\\
(s^*/y)_{AB} &=&-(y/s^*)_{AB}= (s^*p\sigma)_{1}^{AB}g_{2}({\bf k}), \nonumber\\
(s^*/z)_{AB} &=&-(z/s^*)_{AB}= (s^*p\sigma)_{1}^{AB}g_{8}({\bf k}). \nonumber
\end{eqnarray*}
\end{widetext}
where the $g_{ij}({\bf k})$ functions are given, respectively, for 
Si~($111$) and silicene in Eqs.~(\ref{eq:g-func-Si-111}) 
and~(\ref{eq:g-func-graph}).
\begin{widetext}
\begin{eqnarray}\label{eq:g-func-Si-111}
g_{0}({\bf k}) &=& e^{i{\bf k}\cdot{\mbox{\boldmath{$\delta$}}_{1}^{(1)}}}
                + e^{i{\bf k}\cdot{\mbox{\boldmath{$\delta$}}_{2}^{(1)}}}
                + e^{i{\bf k}\cdot{\mbox{\boldmath{$\delta$}}_{3}^{(1)}}},\nonumber\\
g_{1}({\bf k}) &=& -\frac{2 \sqrt{2}}{9}(e^{i{\bf k}\cdot{\mbox{\boldmath{$\delta$}}_{1}^{(1)}}}
                -\frac{1}{2}e^{i{\bf k}\cdot{\mbox{\boldmath{$\delta$}}_{2}^{(1)}}}
                -\frac{1}{2}\,e^{i{\bf k}\cdot{\mbox{\boldmath{$\delta$}}_{3}^{(1)}}}),\nonumber\\
g_{2}({\bf k}) &=& \sqrt{\frac{2}{3}}\left(e^{i{\bf k}\cdot{\mbox{\boldmath{$\delta$}}_{2}^{(1)}}}
                -e^{i{\bf k}\cdot{\mbox{\boldmath{$\delta$}}_{3}^{(1)}}}\right), \nonumber\\
g_{3}({\bf k}) &=& \frac{8}{9}\left(e^{i{\bf k}\cdot{\mbox{\boldmath{$\delta$}}_{1}^{(1)}}} 
                +\frac{1}{4}\,e^{i{\bf k}\cdot{\mbox{\boldmath{$\delta$}}_{2}^{(1)}}}
                 \frac{1}{4}\,e^{i{\bf k}\cdot{\mbox{\boldmath{$\delta$}}_{3}^{(1)}}}\right),\nonumber\\
g_{4}({\bf k}) &=&\frac{1}{9} \left(e^{i{\bf k}\cdot{\mbox{\boldmath{$\delta$}}_{1}^{(1)}}} 
                +7\,e^{i{\bf k}\cdot{\mbox{\boldmath{$\delta$}}_{2}^{(1)}}} 
                +7\,e^{i{\bf k}\cdot{\mbox{\boldmath{$\delta$}}_{3}^{(1)}}}\right),\nonumber\\
g_{5}({\bf k}) &=&-\frac{2}{3}\left(e^{i{\bf k}\cdot{\mbox{\boldmath{$\delta$}}_{2}^{(1)}}}
                +e^{i{\bf k}\cdot{\mbox{\boldmath{$\delta$}}_{3}^{(1)}}}\right), \nonumber\\
g_{6}({\bf k}) &=& \frac{2}{3}\left(e^{i{\bf k}\cdot{\mbox{\boldmath{$\delta$}}_{2}^{(1)}}} 
                +e^{i{\bf k}\cdot{\mbox{\boldmath{$\delta$}}_{3}^{(1)}}}\right),\\
g_{7}({\bf k}) &=& e^{i{\bf k}\cdot{\mbox{\boldmath{$\delta$}}_{1}^{(1)}}}
                + \frac{1}{3}\,e^{i{\bf k}\cdot{\mbox{\boldmath{$\delta$}}_{2}^{(1)}}}
                + \frac{1}{3}\,e^{i{\bf k}\cdot{\mbox{\boldmath{$\delta$}}_{3}^{(1)}}},\nonumber\\
g_{8}({\bf k}) &=&-\frac{1}{3}\left(e^{i{\bf k}\cdot{\mbox{\boldmath{$\delta$}}_{1}^{(1)}}}
                + e^{i{\bf k}\cdot{\mbox{\boldmath{$\delta$}}_{2}^{(1)}}}
                + e^{i{\bf k}\cdot{\mbox{\boldmath{$\delta$}}_{3}^{(1)}}}\right),\nonumber\\
g_{9}({\bf k}) &=& -\frac{2 \sqrt{2}}{9}\left(e^{i{\bf k}\cdot{\mbox{\boldmath{$\delta$}}_{1}^{(1)}}}
                -\frac{1}{2}e^{i{\bf k}\cdot{\mbox{\boldmath{$\delta$}}_{2}^{(1)}}}
                -\frac{1}{2}\,e^{i{\bf k}\cdot{\mbox{\boldmath{$\delta$}}_{3}^{(1)}}}\right),\nonumber\\
g_{10}({\bf k})&=& \frac{1}{3}\sqrt{\frac{2}{3}}\left(e^{i{\bf k}\cdot{\mbox{\boldmath{$\delta$}}_{2}^{(1)}}}
                -e^{i{\bf k}\cdot{\mbox{\boldmath{$\delta$}}_{3}^{(1)}}}\right),\nonumber\\
g_{11}({\bf k})&=&\frac{1}{9}\left(e^{i{\bf k}\cdot{\mbox{\boldmath{$\delta$}}_{1}^{(1)}}}
                + e^{i{\bf k}\cdot{\mbox{\boldmath{$\delta$}}_{2}^{(1)}}}
                + e^{i{\bf k}\cdot{\mbox{\boldmath{$\delta$}}_{3}^{(1)}}}\right),\nonumber\\
g_{12}({\bf k})&=& \frac{8}{9}\left(e^{i{\bf k}\cdot{\mbox{\boldmath{$\delta$}}_{1}^{(1)}}}
                + e^{i{\bf k}\cdot{\mbox{\boldmath{$\delta$}}_{2}^{(1)}}}
                + e^{i{\bf k}\cdot{\mbox{\boldmath{$\delta$}}_{3}^{(1)}}}\right)
                =\frac{8}{9}\left(e^{i\frac{k_{x}a}{\sqrt{3}}}
                +2e^{-i\frac{k_{x}a}{2\sqrt{3}}}\cos\frac{k_{y}a}{2}\right),\nonumber\\
g_{13}({\bf k}) &=& e^{i{\bf k}\cdot{\mbox{\boldmath{$\delta$}}_{1}^{(2)}}}
                + e^{i{\bf k}\cdot{\mbox{\boldmath{$\delta$}}_{2}^{(2)}}}
                + e^{i{\bf k}\cdot{\mbox{\boldmath{$\delta$}}_{3}^{(2)}}}
                + e^{i{\bf k}\cdot{\mbox{\boldmath{$\delta$}}_{4}^{(2)}}}
                + e^{i{\bf k}\cdot{\mbox{\boldmath{$\delta$}}_{5}^{(2)}}}
                + e^{i{\bf k}\cdot{\mbox{\boldmath{$\delta$}}_{6}^{(2)}}},\nonumber\\
g_{14}({\bf k}) &=&\frac{\sqrt{3}}{2}\left(e^{i{\bf k}\cdot{\mbox{\boldmath{$\delta$}}_{3}^{(2)}}}
                -e^{i{\bf k}\cdot{\mbox{\boldmath{$\delta$}}_{4}^{(2)}}}
                +e^{i{\bf k}\cdot{\mbox{\boldmath{$\delta$}}_{5}^{(2)}}}
                -e^{i{\bf k}\cdot{\mbox{\boldmath{$\delta$}}_{6}^{(2)}}}\right), \nonumber\\
g_{15}({\bf k}) &=& e^{i{\bf k}\cdot{\mbox{\boldmath{$\delta$}}_{1}^{(2)}}}
                + e^{i{\bf k}\cdot{\mbox{\boldmath{$\delta$}}_{2}^{(2)}}}
                +\frac{1}{4}\,e^{i{\bf k}\cdot{\mbox{\boldmath{$\delta$}}_{3}^{(2)}}}
                +\frac{1}{4}\,e^{i{\bf k}\cdot{\mbox{\boldmath{$\delta$}}_{4}^{(2)}}}
                +\frac{1}{4}\,e^{i{\bf k}\cdot{\mbox{\boldmath{$\delta$}}_{5}^{(2)}}}
                -\frac{1}{4}\,e^{i{\bf k}\cdot{\mbox{\boldmath{$\delta$}}_{6}^{(2)}}},\nonumber\\
g_{16}({\bf k}) &=& \frac{3}{4}\left( e^{i{\bf k}\cdot{\mbox{\boldmath{$\delta$}}_{3}^{(2)}}}
                +e^{i{\bf k}\cdot{\mbox{\boldmath{$\delta$}}_{4}^{(2)}}}
                +e^{i{\bf k}\cdot{\mbox{\boldmath{$\delta$}}_{5}^{(2)}}}
                +e^{i{\bf k}\cdot{\mbox{\boldmath{$\delta$}}_{6}^{(2)}}}\right), \nonumber\\
g_{17}({\bf k}) &=& e^{i{\bf k}\cdot{\mbox{\boldmath{$\delta$}}_{1}^{(2)}}} 
                +e^{i{\bf k}\cdot{\mbox{\boldmath{$\delta$}}_{2}^{(2)}}} 
                +\frac{3}{4}\,e^{i{\bf k}\cdot{\mbox{\boldmath{$\delta$}}_{3}^{(2)}}}
                +\frac{3}{4}\,e^{i{\bf k}\cdot{\mbox{\boldmath{$\delta$}}_{4}^{(2)}}} 
                +\frac{3}{4}\,e^{i{\bf k}\cdot{\mbox{\boldmath{$\delta$}}_{5}^{(2)}}} 
                +\frac{3}{4}\,e^{i{\bf k}\cdot{\mbox{\boldmath{$\delta$}}_{6}^{(2)}}},\nonumber\\
g_{18}({\bf k}) &=&\frac{\sqrt{3}}{4}\left(-e^{i{\bf k}\cdot{\mbox{\boldmath{$\delta$}}_{3}^{(2)}}}
                -e^{i{\bf k}\cdot{\mbox{\boldmath{$\delta$}}_{4}^{(2)}}}
                +e^{i{\bf k}\cdot{\mbox{\boldmath{$\delta$}}_{5}^{(2)}}}
                +e^{i{\bf k}\cdot{\mbox{\boldmath{$\delta$}}_{6}^{(2)}}}\right)\nonumber\\
g_{19}({\bf k}) &=&e^{i{\bf k}\cdot{\mbox{\boldmath{$\delta$}}_{1}^{(2)}}} 
                +e^{i{\bf k}\cdot{\mbox{\boldmath{$\delta$}}_{2}^{(2)}}} 
                +\frac{1}{4}\,e^{i{\bf k}\cdot{\mbox{\boldmath{$\delta$}}_{3}^{(2)}}}
                +\frac{1}{4}\,e^{i{\bf k}\cdot{\mbox{\boldmath{$\delta$}}_{4}^{(2)}}} 
                +\frac{1}{4}\,e^{i{\bf k}\cdot{\mbox{\boldmath{$\delta$}}_{5}^{(2)}}} 
                +\frac{1}{4}\,e^{i{\bf k}\cdot{\mbox{\boldmath{$\delta$}}_{6}^{(2)}}},\nonumber\\
g_{20}({\bf k}) &=& \frac{3}{4}\left( e^{i{\bf k}\cdot{\mbox{\boldmath{$\delta$}}_{3}^{(2)}}}
                +e^{i{\bf k}\cdot{\mbox{\boldmath{$\delta$}}_{4}^{(2)}}}
                +e^{i{\bf k}\cdot{\mbox{\boldmath{$\delta$}}_{5}^{(2)}}}
                +e^{i{\bf k}\cdot{\mbox{\boldmath{$\delta$}}_{6}^{(2)}}}\right), \nonumber\\
g_{21}({\bf k})&=& g_{22}({\bf k})= g_{23}({\bf k})=0,\nonumber\\
g_{24}({\bf k})&=&  e^{i{\bf k}\cdot{\mbox{\boldmath{$\delta$}}_{1}^{(2)}}}
                + e^{i{\bf k}\cdot{\mbox{\boldmath{$\delta$}}_{2}^{(2)}}}
                + e^{i{\bf k}\cdot{\mbox{\boldmath{$\delta$}}_{3}^{(2)}}}
                + e^{i{\bf k}\cdot{\mbox{\boldmath{$\delta$}}_{4}^{(2)}}}
                + e^{i{\bf k}\cdot{\mbox{\boldmath{$\delta$}}_{5}^{(2)}}}
                + e^{i{\bf k}\cdot{\mbox{\boldmath{$\delta$}}_{6}^{(2)}}},\nonumber
\end{eqnarray}
\end{widetext}    
where the position vectors 
$\mbox{\boldmath{$\delta$}}_{j}^{(k)}$ 
are given in Eqs.~(\ref{eq:Si(111)_1NN}) 
and (\ref{eq:Si(111)_2NN}).\\  
\begin{widetext}
\begin{eqnarray}\label{eq:g-func-graph}
g_{0}({\bf k}) &=& e^{i{\bf k}\cdot{\mbox{\boldmath{$\delta$}}_{1}^{(1)}}}
                + e^{i{\bf k}\cdot{\mbox{\boldmath{$\delta$}}_{2}^{(1)}}}
                + e^{i{\bf k}\cdot{\mbox{\boldmath{$\delta$}}_{3}^{(1)}}},\nonumber\\
g_{1}({\bf k}) &=& e^{i{\bf k}\cdot{\mbox{\boldmath{$\delta$}}_{1}^{(1)}}}
                -\frac{1}{2}e^{i{\bf k}\cdot{\mbox{\boldmath{$\delta$}}_{2}^{(1)}}}
                -\frac{1}{2}\,e^{i{\bf k}\cdot{\mbox{\boldmath{$\delta$}}_{3}^{(1)}}},\nonumber\\
g_{2}({\bf k}) &=& \frac{\sqrt{3}}{2}\left(e^{i{\bf k}\cdot{\mbox{\boldmath{$\delta$}}_{2}^{(1)}}}
                -e^{i{\bf k}\cdot{\mbox{\boldmath{$\delta$}}_{3}^{(1)}}}\right), \nonumber\\
g_{3}({\bf k}) &=& e^{i{\bf k}\cdot{\mbox{\boldmath{$\delta$}}_{1}^{(1)}}} 
                +\frac{1}{4}\,e^{i{\bf k}\cdot{\mbox{\boldmath{$\delta$}}_{2}^{(1)}}}
                 \frac{1}{4}\,e^{i{\bf k}\cdot{\mbox{\boldmath{$\delta$}}_{3}^{(1)}}},\nonumber\\
g_{4}({\bf k}) &=&\frac{3}{4}\left( 
                 e^{i{\bf k}\cdot{\mbox{\boldmath{$\delta$}}_{2}^{(1)}}} 
                +e^{i{\bf k}\cdot{\mbox{\boldmath{$\delta$}}_{3}^{(1)}}}\right),\nonumber\\
g_{5}({\bf k}) &=&-\frac{\sqrt{3}}{4}\left(e^{i{\bf k}\cdot{\mbox{\boldmath{$\delta$}}_{2}^{(1)}}}
                -e^{i{\bf k}\cdot{\mbox{\boldmath{$\delta$}}_{3}^{(1)}}}\right), \nonumber\\
g_{6}({\bf k}) &=& \frac{3}{4}\left(e^{i{\bf k}\cdot{\mbox{\boldmath{$\delta$}}_{2}^{(1)}}} 
                +e^{i{\bf k}\cdot{\mbox{\boldmath{$\delta$}}_{3}^{(1)}}}\right), \nonumber\\
g_{7}({\bf k}) &=& e^{i{\bf k}\cdot{\mbox{\boldmath{$\delta$}}_{1}^{(1)}}}
                + \frac{1}{4}\,e^{i{\bf k}\cdot{\mbox{\boldmath{$\delta$}}_{2}^{(1)}}}
                + \frac{1}{4}\,e^{i{\bf k}\cdot{\mbox{\boldmath{$\delta$}}_{3}^{(1)}}},\nonumber\\
g_{8}({\bf k}) &=& g_{9}({\bf k})= g_{10}({\bf k}) = g_{11}({\bf k})= 0,\\
g_{12}({\bf k})&=& e^{i{\bf k}\cdot{\mbox{\boldmath{$\delta$}}_{1}^{(1)}}}
                + e^{i{\bf k}\cdot{\mbox{\boldmath{$\delta$}}_{2}^{(1)}}}
                + e^{i{\bf k}\cdot{\mbox{\boldmath{$\delta$}}_{3}^{(1)}}}
                = e^{i\frac{k_{x}a}{\sqrt{3}}}
                + 2e^{-i\frac{k_{x}a}{2\sqrt{3}}}\cos\frac{k_{y}a}{2},\nonumber\\
g_{13}({\bf k}) &=& e^{i{\bf k}\cdot{\mbox{\boldmath{$\delta$}}_{1}^{(2)}}}
                + e^{i{\bf k}\cdot{\mbox{\boldmath{$\delta$}}_{2}^{(2)}}}
                + e^{i{\bf k}\cdot{\mbox{\boldmath{$\delta$}}_{3}^{(2)}}}
                + e^{i{\bf k}\cdot{\mbox{\boldmath{$\delta$}}_{4}^{(2)}}}
                + e^{i{\bf k}\cdot{\mbox{\boldmath{$\delta$}}_{5}^{(2)}}}
                + e^{i{\bf k}\cdot{\mbox{\boldmath{$\delta$}}_{6}^{(2)}}},\nonumber\\
g_{14}({\bf k}) &=&\frac{\sqrt{3}}{2}\left(e^{i{\bf k}\cdot{\mbox{\boldmath{$\delta$}}_{3}^{(2)}}}
                -e^{i{\bf k}\cdot{\mbox{\boldmath{$\delta$}}_{4}^{(2)}}}
                +e^{i{\bf k}\cdot{\mbox{\boldmath{$\delta$}}_{5}^{(2)}}}
                -e^{i{\bf k}\cdot{\mbox{\boldmath{$\delta$}}_{6}^{(2)}}}\right), \nonumber\\
g_{15}({\bf k}) &=& e^{i{\bf k}\cdot{\mbox{\boldmath{$\delta$}}_{1}^{(2)}}}
                - e^{i{\bf k}\cdot{\mbox{\boldmath{$\delta$}}_{2}^{(2)}}}
                -\frac{1}{2}\,e^{i{\bf k}\cdot{\mbox{\boldmath{$\delta$}}_{3}^{(2)}}}
                +\frac{1}{2}\,e^{i{\bf k}\cdot{\mbox{\boldmath{$\delta$}}_{4}^{(2)}}}
                +\frac{1}{2}\,e^{i{\bf k}\cdot{\mbox{\boldmath{$\delta$}}_{5}^{(2)}}}
                -\frac{1}{2}\,e^{i{\bf k}\cdot{\mbox{\boldmath{$\delta$}}_{6}^{(2)}}},\nonumber\\
g_{16}({\bf k}) &=& \frac{3}{4}\left( e^{i{\bf k}\cdot{\mbox{\boldmath{$\delta$}}_{3}^{(2)}}}
                +e^{i{\bf k}\cdot{\mbox{\boldmath{$\delta$}}_{4}^{(2)}}}
                +e^{i{\bf k}\cdot{\mbox{\boldmath{$\delta$}}_{5}^{(2)}}}
                +e^{i{\bf k}\cdot{\mbox{\boldmath{$\delta$}}_{6}^{(2)}}}\right), \nonumber\\
g_{17}({\bf k}) &=& e^{i{\bf k}\cdot{\mbox{\boldmath{$\delta$}}_{1}^{(2)}}} 
                +e^{i{\bf k}\cdot{\mbox{\boldmath{$\delta$}}_{2}^{(2)}}} 
                +\frac{1}{4}\,e^{i{\bf k}\cdot{\mbox{\boldmath{$\delta$}}_{3}^{(2)}}}
               +\frac{1}{4}\,e^{i{\bf k}\cdot{\mbox{\boldmath{$\delta$}}_{4}^{(2)}}} 
                +\frac{1}{4}\,e^{i{\bf k}\cdot{\mbox{\boldmath{$\delta$}}_{5}^{(2)}}} 
                +\frac{1}{4}\,e^{i{\bf k}\cdot{\mbox{\boldmath{$\delta$}}_{6}^{(2)}}},\nonumber\\
g_{18}({\bf k}) &=&\frac{\sqrt{3}}{4}\left(-e^{i{\bf k}\cdot{\mbox{\boldmath{$\delta$}}_{3}^{(2)}}}
                -e^{i{\bf k}\cdot{\mbox{\boldmath{$\delta$}}_{4}^{(2)}}}
                +e^{i{\bf k}\cdot{\mbox{\boldmath{$\delta$}}_{5}^{(2)}}}
                +e^{i{\bf k}\cdot{\mbox{\boldmath{$\delta$}}_{6}^{(2)}}}\right),\nonumber\\
g_{19}({\bf k}) &=&e^{i{\bf k}\cdot{\mbox{\boldmath{$\delta$}}_{1}^{(2)}}} 
                +e^{i{\bf k}\cdot{\mbox{\boldmath{$\delta$}}_{2}^{(2)}}} 
                +\frac{1}{4}\,e^{i{\bf k}\cdot{\mbox{\boldmath{$\delta$}}_{3}^{(2)}}}
                +\frac{1}{4}\,e^{i{\bf k}\cdot{\mbox{\boldmath{$\delta$}}_{4}^{(2)}}} 
                +\frac{1}{4}\,e^{i{\bf k}\cdot{\mbox{\boldmath{$\delta$}}_{5}^{(2)}}} 
                +\frac{1}{4}\,e^{i{\bf k}\cdot{\mbox{\boldmath{$\delta$}}_{6}^{(2)}}},\nonumber\\
g_{20}({\bf k})  &=& \frac{3}{4}\left( e^{i{\bf k}\cdot{\mbox{\boldmath{$\delta$}}_{3}^{(2)}}}
                +e^{i{\bf k}\cdot{\mbox{\boldmath{$\delta$}}_{4}^{(2)}}}
                +e^{i{\bf k}\cdot{\mbox{\boldmath{$\delta$}}_{5}^{(2)}}}
                +e^{i{\bf k}\cdot{\mbox{\boldmath{$\delta$}}_{6}^{(2)}}}\right), \nonumber\\
g_{21}({\bf k}) &=& g_{22}({\bf k})= g_{23}({\bf k})=g_{24}({\bf k})=0,\nonumber\\
g_{25}({\bf k})&=&  e^{i{\bf k}\cdot{\mbox{\boldmath{$\delta$}}_{1}^{(2)}}}
                + e^{i{\bf k}\cdot{\mbox{\boldmath{$\delta$}}_{2}^{(2)}}}
                + e^{i{\bf k}\cdot{\mbox{\boldmath{$\delta$}}_{3}^{(2)}}}
                + e^{i{\bf k}\cdot{\mbox{\boldmath{$\delta$}}_{4}^{(2)}}}
                + e^{i{\bf k}\cdot{\mbox{\boldmath{$\delta$}}_{5}^{(2)}}}
                + e^{i{\bf k}\cdot{\mbox{\boldmath{$\delta$}}_{6}^{(2)}}}.\nonumber
\end{eqnarray}
\end{widetext}

\end{document}